# Time-resolved X-ray microscopy of nanoparticle aggregates under oscillatory shear

G.K. Auernhammer[1*], K. Fauth[2,3], B. Ullrich[1], J. Zhao[1], M. Weigand[2], D. Vollmer[1]

[1]*MPI für Polymerforschung, Ackermannweg 10, 55128 Mainz, Germany*
[2]*MPI für Metallforschung, Heisenbergstrasse 3,70569 Stuttgart Stuttgart, Germany*
[3]*Present address: Experimentelle Physik 4, Physikalisches Institut, Am Hubland, 97074 Würzburg, Germany*
[*]*Email: guenter.auernhammer@mpip-mainz.mpg.de*

**Abstract**

Of all current detection techniques with nanometer resolution, only X-ray microscopy allows imaging nanoparticles in suspension. Can it also be used to investigate structural dynamics? When studying response to mechanical stimuli, the challenge lies in applying them with precision comparable to spatial resolution. In the first shear experiments performed in an X-ray microscope, we accomplished this by inserting a piezo actuator driven shear cell into the focal plane of a scanning transmission X-ray microscope (STXM). Thus shear-induced reorganization of magnetite nanoparticle aggregates could be demonstrated in suspension. As X-ray microscopy proves suitable for studying structural change, new prospects open up in physics at small length scales.

**Keywords:** nanoparticles; polymers, shear; microscopy; crosslinking

A number of experimental methods and devices have been developed for space and time-resolved investigation of structural response to mechanical shear, electric fields, temperature and so on in a variety of tuneable environments. So far our knowledge on structure in materials composed of nanometer-sized entities is largely based on electron microscopy and scattering methods. Since the former requires frozen or dried samples and the latter average over large sample volumes, no information on structural changes in solution, the native environment for most nanoparticle systems, can be obtained.

Scanning transmission X-ray microscopes (STXMs) focalize a highly monochromatic synchrotron X-ray beam with the help of Fresnel zone plates. Imaging of the sample is provided by scanning the focal spot over the sample and measuring the transmission point by point. STXMs offer the possibility to resolve individual particles of a wide variety of chemical compositions and sizes down to a few tens of nanometers, i.e. one order of magnitude below those resolved by laser scanning confocal microscopes. Contrast in STXM derives either from the spatial dependence of electron density or from element specific resonances at absorption edges. High quality Fresnel zone plates provide lateral resolution below 35 nm (1). Individual particles of a wide variety of chemical compositions (2) have been resolved with that resolution. The depth of focus in STXM extends over few microns, hence both bulk structure and interface effects can be analyzed, either in transmission or tomography images, as done in biological samples such as fossil embryos (4). STXM has been combined with external magnetic fields (3). Yet the main advantage of X-ray microscopy is that it can be carried out in liquid media, as demonstrated by detection of intracellular flow using contrast-rich nanoparticles (5). Since particle behaviour in solution is strongly influenced by solvent properties, STXM, unlike electron microscopy and scattering methods, offers the hitherto unique possibility to observe structures while modulating solvent-mediated interactions (3).

The time required for image acquisition still pose a serious hindrance for the investigation of dynamical processes such as reorganisation, in particular because of radiation-induced damage to the sample. Especially in polymer melts X-ray irradiation causes changes in the chemical structure and cross-linking (6). Moreover, any external manipulation has to be applied with accuracy at least matching that of the imaging device. Thus, simultaneous imaging and stimulation at nanometer resolution remains a technical challenge.

To explore whether real-time analysis of structural change can be performed using present-day X-ray microscopy, we constructed a shear cell (Fig. 1), to be inserted into the focal plane of the STXM at Paul Scherrer Institute (Villigen, Switzerland). To prevent interfering with the optical components of the microscope, i.e. Fresnel zone plates, pinhole and the detector, the size of our shear cell amounted to 14x18x20 mm$^3$. The sample cell, with a total volume of 10 μl, contained a window consisting of two silicon nitride membranes spaced at about 5 – 10 μm, the regular sample thickness in this device. The oscillatory scanning motion of the sample could be controlled with nm precision. Shear was applied by moving the detector-side cell wall perpendicular to the X-ray beam using a piezo actuator with a maximal displacement of ± 2.5 μm.

As a model system, we chose a suspension of 50 nm sized magnetite particles (7) with attractive effective interactions. Brownian motion and inertia effects were slowed down by dispersing the particles in a polydimethylsiloxane melt (MW$_{PDMS}$ ≈ 18 kg/mol). The particles are known to aggregate and form a space-filling structure at a volume fraction as low as 1% (7). Images of a sample containing about 0.5 vol% magnetite particles (dark) embedded in PDMS (light) are shown in Fig. 2. The electron density contrast proved sufficient to allow measurements far away all absorption edges of the matrix material. At the chosen energy (1000 eV), the dominating contrast mechanism is off-resonant absorption. This high photon energy and low intensities strongly reduced irreversible cross-linking of the polymer matrix.

The particles formed loosely packed aggregates (A) whose size varied between several hundred nm and a few μm. As a transmission method, X-ray microscopy reflects superposition of particles by the greyscale of the image. This implies that the more particles lie along the path of the X-ray beam, the darker the pixel appears. To create a well-defined reference point, we cross-linked the upper left corner of the sample (white outline in Fig. 2) with an X-ray dose approximately two orders of magnitude larger than used for imaging. STXM being a scanning method, acquisition takes approximately 2 min per image. To investigate rearrangements of particles due to Brownian motion, an image was taken after 12 minutes (B), showing hardly any changes. Next we sheared the sample at 200 Hz for 10 sec with 2 μm amplitude. Whereas the cross-linked upper left corner remained almost unchanged, pronounced shear-induced rearrangements of the clusters were visible in the rest of the sample (C). Several clusters broke up, others deformed or combined. Particles moved by several μm. Even after more than ten scans rearrangement, partial disintegration, and growth of aggregates were still observed, indicating that no substantial changes of particle-particle and particle-matrix interactions were induced at the time and energies required. The strong magnetic interaction between particles (exceeding at least ten times the thermal energy) obviously prevented rearrangement at rest as well as complete disintegration during shear. The particles moved while shear was applied and relaxed back to a steady state after discontinuation of shear.

With these experiments, we have shown that it is possible to apply shear to a sample with accuracy and reproducibility comparable to that of X-ray microscopy. By working off-resonant we strongly reduced irreversible cross-linking of the polymer matrix, permitting to investigate the time evolution of particle cluster. Shear-induced reorganization of aggregates of particles a few tens of nanometers in size could be demonstrated by X-ray microscopy even at high shear rates. The successful combination of shear experiments with X-ray microscopy is encouraging since length scales of tens of nanometers become accessible to optical techniques. This opens new perspectives in studying soft matter dynamics. It may soon be possible to quantitatively analyze stimulation-induced changes during the entire stimulation process.

**Acknowledgements**
This work was performed at the Swiss Light Source, Paul Scherrer Institut, Villigen, Switzerland. We are grateful to the machine and beamline groups at PSI whose outstanding efforts have made these experiments possible. DV and GKA acknowledge financial support by the Deutsche Forschungsgemeinschaft through SFB-TR6 and SPP-1273, respectively.

## Figure 1

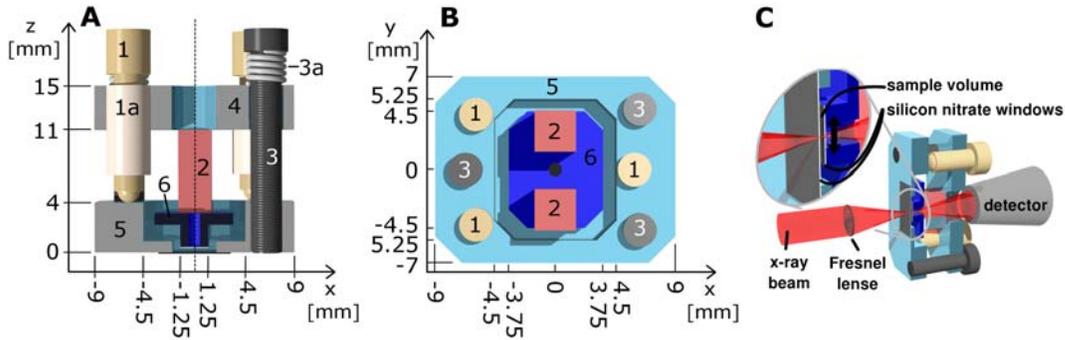

Sketches of the shear cell in coordinate systems with the z-axis along the x-ray beam with z=0 being the sample plane. In panel A a vertical cut through the cell illustrates the design principle, where the dashed line indicates the path of the X-ray beam. The outer part of the shear cell (1, 3, 4, 5) consists of a base plate (4) which holds the front plate (5) via three micrometer screws (1, diameter 2.5 mm, purchased from Owis, Staufen, Germany). The guiding nuts (1a) for the micrometer screws are glued into the base plate with two component adhesive. Both parts are screwed together with three screws (3, diameter 2.5 mm). Springs (3a) between these screws and the base plate allow controlling the force on the micrometer screws. In the inner part of the shear cell (2, 6) two piezo shear actuators (2, purchased from PI Ceramic, Lederhose, Germany) are situated which provide a movement of the inner front plate (6) along the x-axis with an amplitude of max. ± 2.5 µm. The $Si_3N_4$ membrane windows are glued onto the outer (5) and inner (6) front plate with wax. Panel B depicts a cut perpendicular to the beam direction, showing the positioning of the micrometer screws (1), the fixing screws (3), the piezo actuators (2), the outer (5) and inner front plate (6). The black dot sketches the hole for the x-ray beam. Panel C illustrates the position of the shear cell (sample thickness 5-10 µm) in the optical path of the X-ray microscope. Shear is induced by moving one of the two silicon nitride window membranes (inset).

**Figure 2**

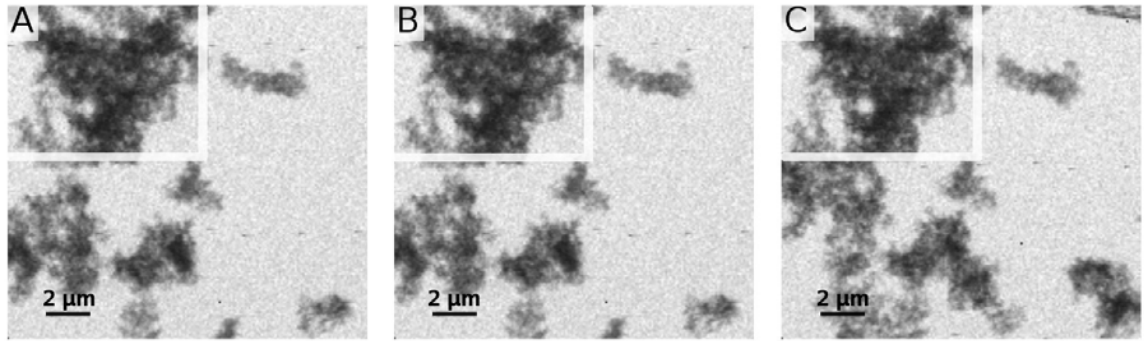

X-ray microscope images taken at 1000 eV showing aggregates of 50 nm sized magnetite particles in a liquid polydimethylsiloxane matrix (A), repeated after 12 min (B), and their reorganization after shear (C). After loading the sample, this sample area has already been scanned 6 times beforehand, demonstrating that even after several scans no significant cross-linking could be detected, i.e. the PDMS matrix is still liquid-like. The area marked in white has been crosslinked prior to the experiment and serves as a reference region.